# *What good is modeling? Introducing ecology and evolution students to theory*

Anna Dornhaus[1], Joanna Masel[1]

[1]Department of Ecology and Evolutionary Biology, University of Arizona, Tucson, AZ, 85721, USA

## Abstract

Theory and empirical science should be in constant dialogue, but often find it hard to understand one another. Here we describe a graduate-level university course we developed to improve matters. The course was designed to help empirically focused biology graduate students read and understand theory papers, despite little prior mathematical training. It uses several evidence-based principles of modern teaching: backwards design, active learning, and just-in-time teaching. This curricular content, emphasizing the nature of evidence and the role of theory in science, aims to improve critical thinking as a means of accelerating scientific progress.

## The problem

There is wide agreement that scientific progress requires integration across empirical and theoretical approaches, including both using theory to interpret and generalize empirical results, and using empirical results to test theories. Mathematical models play a particularly key role in population biology (Provine 1978; Stearns & Schmid-Hempel 1987; Parker & Smith 1990; May 2004; Lander 2010; Gunawardena 2014; Marquet et al. 2014; Servedio et al. 2014; Rosenberg 2020; West et al. 2025).

However, at least three barriers prevent many scientists in ecology, evolutionary biology, epidemiology, and behavior from successfully reading theory papers and understanding their contributions. First, limited mathematical training makes it difficult for many biologists to read theory papers. Biologists may stop reading as soon as equations appear (Fawcett & Higginson 2012) because they worry that they will not be able to follow the thread of the argument (Haller 2014). However, from well-written theory papers, much insight can be obtained even when readers do not follow the mathematical argument itself in detail.

Second, modeling insights have historically come to biological subdisciplines via indirect channels (Provine 1978). Profound insights that originally came from formal models, and which overturned previous conventional wisdom, are often ahistorically passed on to the next generation of scientists as being self-evident from verbal models alone. Even if such a convincing verbal model was subsequently constructed, it does not follow that the scientific community was able to construct it without the help of a previous formal model. A pedagogical emphasis on accessible verbal models to explain past insights can obscure both the historical and current utility of models and theory for biologists.

Third, the role of models in population biology (Servedio et al. 2014) often differs from the role of models in other fields such as physics. See Box 1 for an overview of the powerful scientific method of “Strong Inference”, and the role of models within it. In physics, models generally represent a hypothesis, and are used to derive empirically testable predictions (green #4 in Figure 1). In contrast, in population biology, models are more often used to prove the plausibility of a hypothesis previously thought to be implausible, or to disprove the plausibility of a hypothesis previously thought to be reasonable, without deriving empirically testable predictions (orange #3 in Figure 1; Caswell 1988; Servedio et al. 2014). The dominance of physics as a paradigm for science can lead to profound conceptual misunderstandings regarding what kinds of insights tend to be derived from models in population biology and elsewhere. This can lead to miscommunication and unreasonable expectations from empiricists about what models could or should be able to do. Failures of communication quickly become entrenched when theoreticians and/or empiricists shy away from the difficulties of communicating with one another (Haller 2014).

Compounding this third barrier is that the standard biology curriculum often does not contain any philosophy of science, leaving students uncertain about what constitutes either empirical or theoretical ‘evidence’ for a claim. Understanding what constitutes ‘evidence’ is perhaps the most important element of critical thinking, a skill all education aims to help develop. Rigorous tests of alternative hypotheses, following Strong Inference, are one of the most reliable types of evidence, and yet many researchers in ecology, evolution, and behavior struggle to either articulate clear alternative hypotheses that generate distinctive testable predictions, or to articulate how their work fits into a broader scientific program along these lines (Betts et al. 2021; West et al. 2025). Better understanding of theory can help in these endeavors (Nakagawa et al. 2025).

Our course reconciles an emphasis on alternative hypotheses within Strong Inference with the fact that many important modeling papers treat only a single hypothesis (Caswell 1988; Servedio et al. 2014). The value of these models comes either from showing that a previously proposed verbal hypothesis is internally inconsistent and/or inconsistent with prior knowledge (disproof of principle), or that a previously dismissed verbal hypothesis can in fact be viable (proof of principle). Theoretical work is thus able to adjudicate whether an idea is good enough to even count as a hypothesis worth testing or not, prior to and separate from the process of generating empirically testable predictions. Sometimes at the end of a modeling paper, only one hypothesis (or even zero!) might remain as a viable answer to a scientific question. The more confident the scientific community was concerning a previous verbal hypothesis, the higher impact a theory paper may be that changes its status.

Misunderstandings of the role of theory in the scientific method, and of Strong Inference, can result in absurdities, e.g. insistence that all models be “testable” and that their purpose is to fit data (Ginzburg & Jensen 2004; Epstein 2008). These demands fail to recognize that theory is a powerful tool to examine verbal statements that seem self-evident but may not be.

Defining questions and hypotheses can also be a significant part of the scientific work (#1 and #2 in Figure 1; Betts et al. 2021). Exploration to articulate questions and hypotheses can be seen as more of an art, in contrast to Strong Inference for testing already-articulated hypotheses. Exploration can be inspired by both theory and empirical observations. Indeed, many theoretical projects begin as exploration, and then after unexpected findings are found, pivot to eventually be written up as proof of principle, disproof of principle, or as deriving predictions, as though the question and hypotheses had been there all along.

Finally, and perhaps due to flaws in students' historical and philosophical assumptions, compounded by limited mathematical training (Barraquand et al. 2014), many empirical biologists (from graduate students to professors) have explicitly or implicitly accepted a kind of math phobia. Math phobia is grounded both in the view that models are inaccessible given the empiricist's training, and in the perception that they are ultimately unnecessary to scientific process. We believe both of these views to be mistaken.

------------------------------------------------------------------------------------------------------------

**Box 1: Strong Inference and the role of theory**

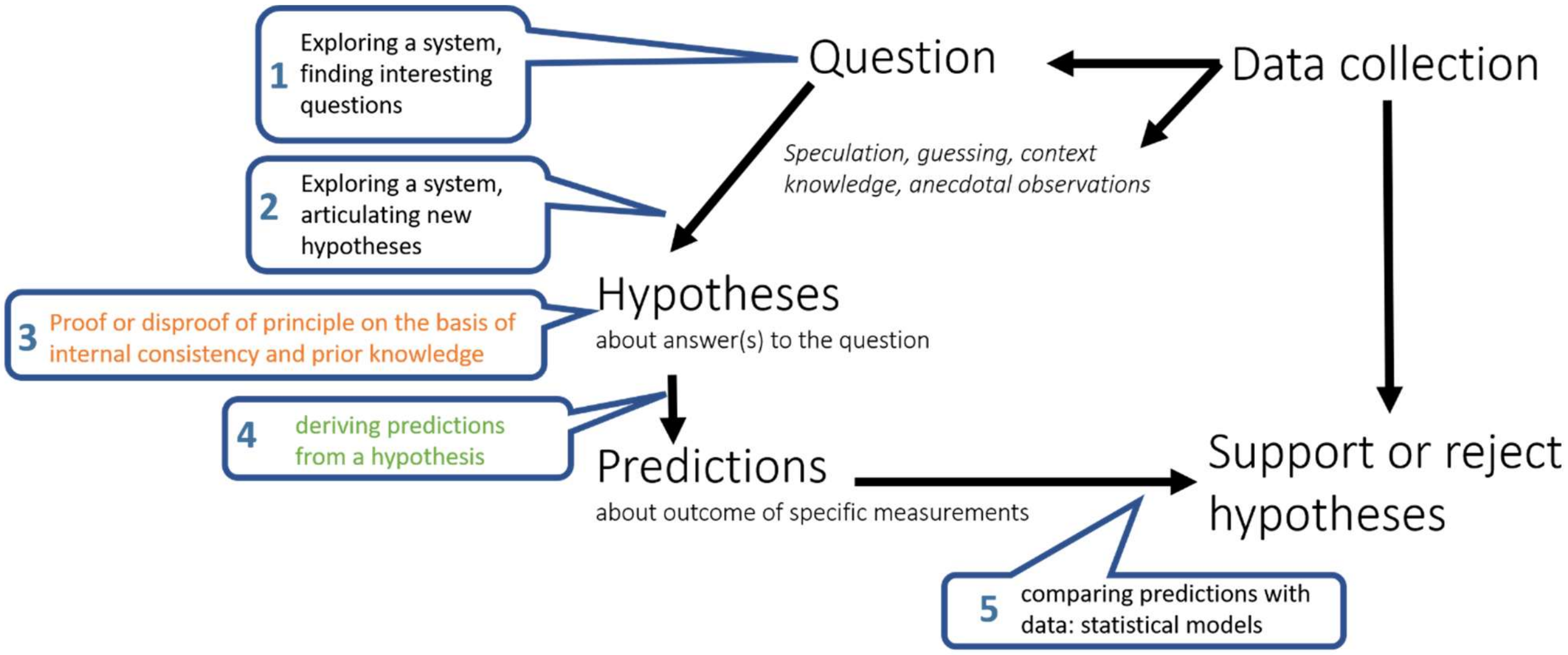

Figure 1: Processes on the left represent contributions of theory (whether verbal or mathematical) to the standard logical flow of Strong Inference, while processes on the right represent those that use data. Our course emphasizes the two uses of models shown in color (3 and 4). We make clear to students that while statistical models are critically important, they fall outside the scope of our class, and that while exploration is part of most modelers' process, it is not generally how a completed paper is written up.

While the nature of scientific approaches is diverse, the structured process of "Strong Inference" (Platt 1964) is generally recognized by scientists as a particularly effective means for arriving at rigorous conclusions (Hutto 2012; Morey et al. 2014; Betini et al. 2017; Nosek et al. 2018; Smith & Cordes 2019; Betts et al. 2021; Nakagawa et al. 2025). Logical deduction can be used to derive specific facts from known general principles; however, the problem of science is typically the reverse, inferring general principles from specific observations. When a specific fact derived ('predicted') from a hypothetical principle is not observed, we can thus infer that the principle (or 'hypothesis') was false. This leads to the 'modus tollens' of Strong Inference: provisional acceptance of a hypothesis is based on the rejection of previously plausible alternatives. Data that agrees with a hypothesis' predictions does not constitute substantive evidence on its own; it must be accompanied by rejection of alternatives. Examining at least two alternative hypotheses is an essential component of the overall pursuit of a scientific question according to these principles (Chamberlin 1890; Platt 1964; Betini et al. 2017). Any conclusion is thus provisional, because novel hypotheses may emerge in future as viable alternatives.

A second major principle of Strong Inference is that all elements of the 'test', i.e. a question, alternative hypotheses, and predictions, are specified prior to knowing the relationship between the data and the predictions (hence '*a priori*'). The reason for this is that motivated reasoning and confirmation bias are powerful forces that may lead authors to use convoluted and varied arguments and modifications of their procedures until a desired result is reached. Specifying the entire procedure of the hypothesis-test, before knowing what its outcome will be, guards against such biased reasoning: this is why many scientists now call for a more formalized process of this specification, such as preregistration (Nosek et al. 2018). This is particularly important since it has become clear that choice of method can have a significant impact on resulting conclusions (Gould et al. 2025). Other strategies used in science similarly aim to prevent researchers from favoring particular results, such as blind observers (Tuyttens et al. 2014; Kardish et al. 2015), standardized statistical procedures (Szucs 2016; Lakens 2021), meta-analyses (Palmer 1999), and many others (McElreath & Smaldino 2015; Fidler et al. 2017).

"Strong Inference" has in the past been framed as "The Scientific Method", and as such invited the criticism that in fact much that is scientific does not involve rejection of hypotheses at all. Actual scientific studies may often use poorly formulated hypotheses that are either not general principles or do not have feasible alternatives (Hutto 2012; Nakagawa et al. 2025). In population biology and other studies of complex real-world systems, outcomes are affected by so many potentially confounded factors that questions of effect size replace mutually exclusive hypotheses about relevant effects (Quinn & Dunham 1983; McElreath & Smaldino 2015). In these cases, Strong Inference is not always the appropriate method; some fields benefit more from exploratory studies (Nakagawa et al. 2025) or from 'descriptive' studies which quantify parameters of interest, for example effect sizes, without testing alternative hypotheses (Quinn & Dunham 1983). However, in any kind of intervention study, as is standard in lab work, the effect size is the direct or indirect result of the intervention, at the cost of not necessarily representing the magnitude of an effect under natural conditions. Thus, the purpose of such studies can only be to actually reject hypotheses.

We include a discussion of the merits of "Strong Inference" in the beginning of our course to ensure students have a broad background in the nature of scientific evidence, prior to more specifically discussing the strength and types of evidence provided by theory.

---------------------------------------------------------------------------------------------------

**Our solution**

To improve communication between empiricists and modelers, and thereby the quality of insights from both, we developed a semester-long course. Our two primary course-level learning objectives are to enable Ph.D. students in ecology and evolution to

1) effectively extract insights for their field from a theory paper even while skipping any math that they cannot follow (for non-theoretically-oriented students), and
2) understand the diverse roles theory plays in generating or testing hypotheses within population biology, and thus how models may improve insights in the field.

We designed the course based on the following evidence-based teaching principles to maximize its impact.

*Backwards design*

Backwards course design (Wiggins & McTighe 2005) is a method to ensure focus on course-level learning objectives while improving teaching effectiveness and simplifying course design. It starts from the 'end', namely the final assessment. This should be designed to test whether course learning goals are reached. Here, since we aimed for students to be able to read a theory paper and extract biological insights and the role of theory in the process, we use a take-home exam in which students are given a theory paper and asked to answer questions about its insights and the role of the models in it. See Supplementary Materials for an example exam.

The course activities are developed only after the nature of the final assessment has been defined. Since "transfer" of knowledge to even mildly different contexts is notoriously difficult (Barnett & Ceci 2002), a recommended approach to achieving specific learning goals is to design the course such that students practice, with feedback, the exact skills needed on as close as possible to the exact same type of problem as represented by the learning goal itself (Banerjee et al. 2025). Thus, to prepare for this final assessment, students read theory papers and answer questions about them throughout the semester (Goudsouzian et al. 2023). For each paper, there is a short in-class quiz, whose question varies relatively little among papers. We ask students to summarize the main scientific contribution of each paper / paper's model, and to classify the model using the scheme in Figure 1 and/or other classification schemes such as deterministic vs. stochastic that are taught in class (see below).

*Active learning*

The paradigm of active learning (Freeman et al. 2014) centers on the idea that students must attempt to solve problems themselves, learn skills and facts in the process, and construct their own knowledge, instead of being passively exposed to information. We are transparent with students about the course learning goals from the first day of class, and encourage them to think of each paper as a practice run both for the final exam and for their own reading of theory papers in their field.

To help students read each paper before class, we provide them with a reading guide, consisting of a set of questions customized to each paper (see Supplementary Materials for the specific questions used for the papers in Table 1). These questions provide scaffolding to prime relevant background knowledge, to direct attention to key parts of the papers needed to contextualize the whole, and to focus students' attention on the two main goals: extract the biological insight(s) from the paper, and define the purpose of the model for achieving these insights. We note that if different papers are chosen, the workload for switching is relatively low – the main task is to generate a comparable set of questions.

We use a variety of conceptual tools (Grainger et al. 2022) to help students classify each paper. One tool that helps students "skip the math" is to identify the inputs and outputs of a model, while treating the model itself as a black box. Another is to classify models using categories such as analytical vs. numerical/simulation, deterministic vs. stochastic, discrete time and/or state vs. continuous time and/or state, well-mixed vs. spatially resolved, quantitative vs. qualitative conclusions etc. These classifications introduce key mathematical concepts at a high level, without requiring students to acquire the technical skills needed to follow the details.

During class time, we discuss the answers to these questions. We use primarily a think-pair-share format, i.e. students write down their answer attempts, then pair up with a neighbor to compare and discuss their answers, before we finally ask them to share answers with the whole group (Smith et al. 2009). While students discuss in pairs, we (the instructors) listen in to various groups to get an early idea of misconceptions or where students may be stuck. This active learning approach gives students ownership of their activity and their learning, and a sense that any explicit 'teaching' they receive serves to help them solve a question they already have (see next section).

*Just-in-time teaching*

Students learn key mathematical concepts in a just-in-time manner, meaning just prior to and/or immediately after they encounter them as an obstacle to their goal of answering questions about a paper. It is critical to teach any ideas, but particularly abstract mathematical concepts, embedded in the context in which we would like students to use them (National Research Council 2003; Matthews et al. 2010; Hester et al. 2014; O'Leary et al. 2021; Garfinkel et al. 2022; Banerjee et al. 2025). We endeavor to enable students to read theory papers without extensive math skills. Nonetheless, we find it is necessary for students to construct a mental representation (Lira & Stieff 2026) for the kinds of mathematical objects used in each paper, even if they do not have the technical ability to manipulate those objects. E.g., students need to know that differential equations can describe continuous-time change in a system over time, and that setting the rate of change to zero can be used to determine a system's equilibrium state.

In addition to using short, just-in-time lecture segments in the class, we also teach using the programming language Mathematica, and give students assignments using it. We have found, supported by post-class student feedback, that programming assignments reinforce students' emerging mental representations of unfamiliar kinds of mathematical objects. For example, we may ask students to reproduce carefully selected figures from the papers they read, or other components of the models. Mathematica enables students (and us) to manipulate equations symbolically and to derive analytical solutions with only high-level understanding of the mathematics. These activities reinforce a high-level, but nonetheless concrete, understanding of what the mathematical objects within the models are, and what can be extracted from them, while skipping the technicalities of how mathematical solutions are derived. We include both a PDF version and the Mathematica 'notebook' versions we use in class, in the Supplementary Materials. In class, we initially distribute a pared-down version of these notebooks, and develop the material and do the exercises with or in front of students. Only after class do we provide the full notebook.

*Team teaching*

Our course was co-developed by a (primarily) theoretician (JM) and a (primarily) empiricist (AD), and we generally also teach it together. Unlike some team teaching, we are both present for almost every class. This enables the students to see the two of us interact and answer each other's questions. The empiricist instructor may notice that some concept is discussed in a way that non-theory-minded students struggle with, and the theoretician instructor may be better able to resolve any issues with mathematical understanding.

Beyond the specific combination of instructor expertise in our course, conversation between instructors can help students see these instructors develop ideas and solve questions in real time, a process also supported by the fact that we (re)develop solutions to most of the Mathematica notebook exercises in front of students. Watching experts 'think aloud' and do worked examples are key evidence-based teaching strategies (e.g. Atkinson et al. 2000; Tang et al. 2025); both can be implemented by a solo instructor, but are facilitated by teaching as a team.

Our course development process involved instructors publicly disagreeing with one another and having non-scripted discussions in front of students. While we believe this to be positive for graduate student education, we note that some students disliked watching professors disagree – surprisingly, including for a discussion that we, the instructors, had thought confirmed that we were in agreement. These discussions tended to be most active in the first few iterations of the course, and were an integral part of course development, until we converged in particular on a shared understanding of the material in Box 1. In other words, in addition to the benefits to students during active instruction, our teamwork and mutual interrogation of content allowed us to refine and clarify our own understanding of 'evidence' and the role of theory. The ongoing conversation between us in and outside of the class enabled us to thoroughly understand different perspectives on the role of modeling in science, in a way that would not have happened had either of us taught a similar course on our own.

**The papers**

We chose the papers (listed in Table 1) to be historically important in making significant advances within their respective fields, and also to represent a diversity of

- the roles theory can play within the scientific method (points 1-4 in Fig. 1)
- mathematical approaches (particularly analytical vs. numerical, deterministic vs. stochastic, temporally dynamic vs. equilibrium)
- subdisciplines of ecology and evolution

The first year we taught the course, we focused on papers that were seminal for the research of students who happened to be in the class that year. At the end of each iteration of the course, we polled the students about which papers and activities they found most vs. least valuable; we then dropped some papers and added others. Over time, this iterative process led to a very complementary set over all three axes of diversity. In the context of discussing the scientific method, the students also read two perspective pieces (Servedio et al. 2014; Betts et al. 2021).

The papers were chosen with an emphasis on the modeling performed. When we polled colleagues for which papers they would like us to teach, many nominated papers that develop a key piece of mathematical theory, e.g. general frameworks for studying species coexistence conditions or allele frequency trajectories. These tend to be most important to students in the corresponding subdiscipline, but are not necessarily accessible to a broad range of student disciplinary backgrounds. Instead, we chose papers that clearly illustrate how models can generate evidence for a claim (see Box 1, and Table 1, 3rd column), and the variety of modeling strategies that may be used for this (Table 1, 4th column, and bullet points above). In many semesters, we are able to add one more paper discussion towards the end of the course, chosen on the basis of the interests of currently enrolled students. Since students are by then equipped with the tools to think about different theory approaches, this is a chance for them to experience applying the tools to a paper directly relevant for them (or for the majority of the class).

**Table 1. Modeling/theory papers used in the course.**

| Paper | Biology | Scientific Method | Math/Model |
|---|---|---|---|
| (Hardy 1908) | Hardy-Weinberg equilibrium | Disproof of principle | Algebra |
| (Schmid-Hempel et al. 1985) | Optimal foraging | Generate predictions | Algebra |
| (Masel et al. 1999) | Prions, r vs $R_0$ | Failure to disprove principle | System ODEs, exponentials |
| (Maynard Smith & Price 1973) | Evolution of ritualized contests | Proof of principle | ESS, simulation |
| (Williams & Martinez 2000) | Food webs | Proof of principle | Networks |
| (Fussmann & Blasius 2005) | Lotka-Volterra cycles, paradox of enrichment | Proof of principle | Dynamical systems, sensitivity to model structure |
| (Stainforth et al. 2005) | Climate | Forecasting | Parameter uncertainty |
| (Luria & Delbrück 1943) | Spontaneous mutation, fluctuation test | Generate predictions | Probability distributions |
| (Hubbell 1997) | Neutral theory of ecology | Proof of principle | Markov chains |
| (Geritz et al. 1997) | Eco-evolutionary feedbacks | Proof of principle | Adaptive dynamics, graphical argument |

## Key insights from each paper

Hardy 1908: ‘Hardy-Weinberg equilibrium’ or even the ‘Hardy-Weinberg law’ is generally taught as a seminal mathematical theory that makes empirically testable predictions, namely that if genotype frequencies are not equal to $p^2$, $2pq$, and $q^2$, where $p$ and $q$ are allele frequencies, then one of the assumptions of the model must be incorrect. However, the corresponding seminal paper does no such thing. Instead, it uses theory to disprove, without any empirical evidence beyond the general rules of Mendelian inheritance, the previous verbal theory that a dominant allele would become more frequent in the population. The ideas that it proposes instead, that dominance is not the same thing as selection, and that there is nothing magic about the 50% allele frequency that gives rise to a 3:1 phenotype ratio, are now seen as obvious, but were not at the time. Opening with this paper helps "shock" students into seeing the gulf between the taught-to-undergraduates version of the role of theory and its actual roles in scientific practice. Another advantage of teaching this paper early is that it is accessible with only relatively straightforward algebra.

Schmid-Hempel et al. 1985: In an informative contrast with Hardy 1908, this paper both derives and empirically tests predictions about how foraging bees should behave under two different assumptions about what bees are maximizing: rate of return (i.e. net energy collected per time) or efficiency (net energy collected per energy spent). These two alternative assumptions thus function as alternative hypotheses. Again, the paper is accessible with fairly straightforward algebra. In preparation, we teach Charnov's classic paper on the marginal value theorem (Charnov 1976), to expose students both to the mathematical strategy of graphical problem-solving, and to the contextual subject matter of optimal foraging theory.

Masel et al. 1999: This paper, on the mechanism by which prions replicate exponentially, introduces students not just to differential equations, but also to coupled differential equations. This includes the representation of mass action differential equations as a ball-and-stick diagram. It also includes the abstraction that $dy_i/dt$ can represent an infinite number of equations, each of them with a different value of *i*. We contrast the Malthusian parameter *r* with the basic reproductive number $R_t$, including digressions to epidemiology. Building subtlety regarding the roles of theory, we discuss how the model had the potential to act as disproof of principle, but did not do so, which adds to the evidence base of the corresponding hypothesis, to which there was (and still is) no clear alternative. Its increased mathematical complexity (compared to the last paper in the sequence) allows us to build techniques and confidence to extract meaning both from the model overall and even from specific equations, even with very little mathematics background (see our reading guide in Supplementary for details). The Mathematica homework asks students to plug in reasonable parameter values and initial conditions to obtain realistic behavior. This challenges students to realize that it is not sufficient for their guesses to stay within a more human-cognition-accessible range of e.g. 1/1000 to 1000, but that considerations such as Avogadro's number necessitate broader thinking about possibilities.

Maynard Smith and Price 1973: This seminal paper on the evolution of ritualized contest behavior played a pivotal role in introducing game theory to biology, as well as the idea that organisms may refrain from all-out conflict even for selfish (i.e. not group-success-based) reasons. In our sequence, it is the first paper whose major results rely on numerical simulations rather than analytical solutions to equations. We ask students to replicate a part of the simulation by writing their own program. This demands the precision of thought needed to fully specify a complex verbal description (of the time course of a contest and the effect of different strategies) as a computer program, a key skill that some students have not previously been exposed to. Prior to this paper, biologists did not consider individual-level selection to be a viable hypothesis to explain ritualized, limited war, leaving group selection as the only viable hypothesis. This paper proves the principle that individual-level selection is in fact a viable hypothesis, re-opening the question of how limited war evolves.

Williams and Martinez 2000 formalize three different simplified models of how the structure of food webs might be generated, and then use simulations to numerically fit all three to empirical data. This paper introduces students to thinking about networks, and to top-down empirical parameterization of models. The mechanics of the paper resemble a process of generating and testing empirical predictions, with one of the three models fitting the data dramatically better than the other two. However, the authors' stated aim is to demonstrate a principle, namely that a very simple generative model can replicate the structural 'complexity' of empirical food webs. They do not claim that the best of their three simple models in fact describes how food webs are generated. This creates an excellent opportunity in the class to discuss the possibility that the same paper can be interpreted to fit multiple categories, and then to further discuss the dominance of proof of principle in population biology models.

Fussmann and Blasius 2005: This paper is a meta-reflection on how subtle choices of functional form might affect conclusions drawn from a model, a consideration that is often surprising to the empirical biologist. The paper demonstrates this by analyzing the stability of equilibria in an ecological dynamical systems model. We therefore preface discussion of this paper with a high-level introduction to dynamical systems, using Mathematica to visualize trajectories in two-dimensional phase space. We use a one-dimensional graphical argument to motivate the use of second derivatives to assess the stability of equilibria, and then build on the matrix work of the previous paper to generalize to a Jacobian matrix. We apply these tools to models within the paper itself to recapitulate the authors' insights.

Stainforth et al. 2005: We classify this climate science paper as outside the scope of the scientific method of Figure 1. We instead see it, together with all climate models, as "forecasting" climate trajectories under different scenarios, which is distinct from making empirically testable predictions for the purpose of supporting one hypothesis over another. In applied sciences / engineering, forecasts are interesting insights in and of themselves, without being part of a process of hypothesis-testing. This is similarly true e.g. for demographic forecasting, whether for human demography or for wildlife biology. The goal of forecasting is to inform decision-making, in contrast to 'explanatory models', which derive conceptual insights (Sarewitz & Pielke 1999; Mouquet et al. 2015). We chose this paper rather than a more up-to-date climate forecast because its focus is on expanding the range of uncertainties considered. Previous climate models quantified only the uncertainty arising from a chaotic dynamical system, via perturbations to initial conditions. This paper was the first to also explore uncertainty in parameter values.

Luria and Delbrück 1943: This seminal paper develops testable predictions for the alternative hypotheses for how phage-resistance mutations appear. The induced mutation hypothesis predicts a Poisson distribution where the variance across replicate experiments equals the mean, while the spontaneous mutation hypothesis predicts variance far greater than the mean. This puts the emphasis squarely on probability distributions as a mathematical object of study, and so we first introduce probability distributions (not just the Poisson).

Hubbell 1997: Following our coverage of probability distributions, we are next able to introduce transition matrices and finite Markov chains. Students often believe that stochastic systems can only be tackled through simulations, and are surprised to learn about analytical approaches. Mathematica can greatly increase the accessibility of solution techniques, allowing students to focus on understanding the nature of the mathematical objects. In this new and more challenging context, students again tend to struggle with the idea that equations tracking the probability of being in state $N_i$ represent an infinite number of equations, each of them with a different value of $i$. We also include substantial discussion of the sometimes-contentious role of the neutral theory of ecology, first introduced in this paper, in the scientific method. We frame the paper's contribution as showing that certain types of data cannot be used to draw detailed conclusions, because the paper proves the principle that even a simplified, obviously wrong model is able to predict these forms of data.

Geritz et al. 1997: We use this adaptive dynamics paper to introduce graphical arguments, and the beauty that more mathematically-oriented theoreticians find in generality. It proves the principle that eco-evolutionary dynamics can create branching points, which can be interpreted as speciation. The paper also introduces students to the idea of separation of timescales. Because of this paper's more mathematical style, the reading guide is particularly important, helping students build confidence in their ability to read even more abstract papers.

**Challenges for the course**

As mentioned earlier, biologists often broadly accept that there should be integration across empirical science and theory, but struggle to recognize how the 'other' approach can inform their specific research. One of our primary challenges before the first day of class, therefore, is to convince the students whom our course targets, together with their advisors, that students without specific background, mathematical skills, or plans to do any modeling themselves might benefit from, and will be capable of completing, the course. In fact, many of our students have had little math training, and the class is specifically set up to teach students to extract what they need from papers without 'doing the math'.

We unexpectedly found that students who are theoreticians, e.g. in an Applied Math Ph.D. program, not only enroll in our course, but seem to benefit for overlapping reasons: seeing the forest for the trees to understand what modeling is useful for, and how it interacts with empirical science. Our class also attracts some undergraduates, usually with above-average mathematical preparation, experience of biology research, and intent to enroll in a Ph.D. program. It has occasionally also attracted students from other branches of biology, e.g. molecular biology, and neuroscience.

Enormous diversity among students in their mathematical preparation is thus a possible challenge for our course. However, we have found this to be far less of a problem than it might appear. In our active learning approach, students work at their own level of understanding in pairs, and often teach each other (with benefits to both students). These interactions among budding scientists with different backgrounds and interests reinforce, for both partners, the communication skills needed for interdisciplinary research. A key teaching strategy is to direct student pairs to first let whichever student has less background articulate questions or attempt answers, in order to allow them to make progress in their thinking, before their more experienced partner 'reveals' answers.

Nonetheless, more extreme cases of 'math phobia' or lack of fluency with more elementary mathematical skills can present challenges, such as when students struggle with basic algebra, understanding of probability, or entry-level calculus concepts (what is a slope, what is a derivative). In these cases, individualized attention by instructors is useful to help students catch up to a level where effective discussion is possible. This implies that when teaching a student population with low math preparation, it is useful if the class remains at relatively low enrollment (which has always been in the range of 6-16 students).

Finally, student commitment to investing time in reading papers is critical. Any course benefits from motivated students, but many biologists are used to being able to ‘skim’ empirical papers and get important information very quickly. This will not work for the modeling papers that we assign, even while “skipping the math” in the sense of not following mathematical derivations. We only expect students to arrive at an understanding of the model as an input-output system, with attention to what exactly goes in (assumptions, parameters, etc.) and what conclusion results from the mathematical argument, but even this requires proceeding more slowly than students expect.

Moreover, the active-learning-based format of the course requires students to grapple with the paper and the questions about it before coming to class. To ensure this, we employ a graded quiz for each paper, often but not always with the same two questions: “What is the major contribution of this paper?” and “How does its use of modeling fit within our Figure 1 scheme and why?” Students know the date by which they are expected to have read each paper, although sometimes the quiz will, in an unannounced fashion, be held in a later class period (since the discussion of each paper typically extends over multiple class periods). We do this when we feel students should have the benefit of discussion before answering the quiz question(s). In addition, the reading guides smooth out sticking points that might have otherwise prevented students from extracting meaning from their reading. Student post-class quantitative and qualitative feedback confirms that they perceive little to no “busy work” over the course of the semester, but rather that time spent is directly related to learning objectives.

As the class progresses, we assign increasingly complex papers, while continuing to provide high levels of scaffolding for at-home reading. The final exam paper is then based on a less complex paper that feels easier in comparison, such that students can interpret it with little low-level guidance.

We note that this course has not yet addressed the challenges of LLMs. Our quizzes and final exams could not have been completed by a chatbot given the exam paper alone, without also supplying the framework introduced in Figure 1. However, a contemporary chatbot could complete them if also supplied with sufficient course-specific material as context. Looking forward, instructors may have to modify the type and procedure of assessments to address this issue. For example, a written take-home could be followed by a brief oral in-class exam on its content. However, we note that in our version of the class, the take-home final only contributes a moderate amount to the overall course grade (see syllabus in Supplementary Materials). In addition, throughout the course, we emphasize the applicability of the skills learned in this course to student research later on; the reason for having a final exam at all is primarily to motivate students and clarify what skills, exactly, they should be working toward. The final exam functions as a summative evaluation only as a distant second goal.

### Relationship to AAAS Vision and Change Framework

The AAAS Vision and Change Framework to reform biology education emerged from a year of conversations among biology faculty, students, administrators, and professional societies, culminating in a meeting of more than 500 biologists (Brewer & Smith 2011). The resulting Framework argued for shifting away from covering biology content to nurturing a smaller set of core concepts and scientific competencies.

Our course closely engages with five of the six core competencies that the framework identified: the process of science, quantitative reasoning, modeling and simulation, tapping into the interdisciplinary nature of science, and communication / collaboration. Within the more fine-grained Bioskills Guide (Clemmons et al. 2020) expanding on the same framework, for the process of science, we address program-level learning objectives of scientific thinking, question formulation, and study design. Within quantitative reasoning, we address the program-level learning objective of numeracy. Within modeling, we address all three program-level learning objectives: the purpose of models, model application, and modeling. We address both interdisciplinary program-level learning objectives: connecting scientific knowledge and interdisciplinary problem solving. Finally on the communications front, we particularly emphasize the program-level learning objective of metacognition, while also engaging in collaboration and collegial review.

### Conclusion

The course 'Introduction to Modeling in Biology' as described here has been successfully taught by us approximately every other year at the University of Arizona since 2014, to students of widely varying backgrounds. We expect modified versions of its approach to work well in other scientific fields. The general approach is adaptable to a different set of papers as considered relevant for the discipline and student population. E.g. to make the course more biomedical, or more specialized in a single field such as only ecology or only evolutionary biology, one would only have to choose papers from those fields along the same lines. Indeed, we are about to adapt it for a more biomedical audience, including biochemical cooperativity (Huang & Ferrell 1996), and cancer as a multi-hit process (Armitage & Doll 1957).

Our course fits with science curricular reform that places what constitutes evidence at the center, with particular emphasis on Strong Inference as a reliable way to make scientific progress. This understanding must span both empirical and theoretical forms of evidence, and their interaction within an overarching practice of science. It must reach both theoretically-oriented and empirically-oriented scientists in training. In a world in desperate need of a general public that can assess the evidence for different claims, university education should ensure that such skills are imparted to budding scientists.

**Acknowledgements**

We thank all the students who have taken versions of this class, from whom we have learned a lot. We thank Ulises Hernández and Mary Regelski for helpful comments on the manuscript.